\documentclass[aps,prc,twocolumn,groupedaddress,showpacs,nofootinbib]{revtex4-1}

\usepackage{graphicx}
\usepackage{dcolumn}
\usepackage{bm}
\usepackage{amssymb,amsmath}
\usepackage{longtable}
\usepackage{enumerate}

\def\F{{\cal F}}

\begin{document}

\title{Precise measurement of branching ratios in the $\beta$ decay of $^{38}$Ca}

\author{H. I. Park}
\email[]{hpark@comp.tamu.edu}
\author{J. C. Hardy}
\email[]{hardy@comp.tamu.edu}
\author{V. E. Iacob}
\author{M. Bencomo}
\author{L. Chen}
\author{V. Horvat}
\author{N. Nica}
\author{B. T. Roeder}
\author{E. McCleskey}
\author{R. E. Tribble}
\author{I. S. Towner}
\affiliation{Cyclotron Institute, Texas A$\&$M University, College Station, Texas 77843-3366, USA}

\date{\today}

\begin{abstract}
We present the full description of a measurement of the branching ratios for the $\beta$-decay of $^{38}$Ca.  This
decay includes five allowed $0^+$$\rightarrow$$\,$$1^+$ branches and a superallowed $0^+$$\rightarrow$$\,$$0^+$ one.
With our new result for the latter, we determine its $ft$ value to be 3062.3$\pm$6.8 s, a result whose precision
(0.2\%) is comparable to the precision of the thirteen well known $0^+$$\rightarrow$$\,$$0^+$ transitions used up
till now for the determination of $V_{ud}$, the up-down quark-mixing element of the CKM matrix.  The $^{38}$Ca
superallowed transition thus becomes the first addition to this set of transitions in nearly a decade and the first
for which a precise mirror comparison is possible, thus enabling an improved test of the isospin-symmetry-breaking
corrections required for the extraction of $V_{ud}$.      
\end{abstract}


\maketitle

\section{Introduction}
\label{sec:intro}

The $\beta$ decay of $^{38}$Ca has special interest.  Its strongest branch is a superallowed transition from the $0^+$, $T$=1
ground state of $^{38}$Ca to the $0^+$, $T$=1 analog state at 130-keV excitation energy in its daughter $^{38}$K.  Such
$0^+$$\rightarrow$$\,$$0^+$ transitions have long played an important role in yielding the most precise value for
$V_{ud}$, the up-down quark-mixing element of the Cabibbo-Kobayashi-Maskawa (CKM) matrix, which in turn enables the
most demanding test of that matrix's unitarity.  Although the standard model does not specify values for the CKM matrix
elements, it does require the matrix to be unitary.  If the measured matrix elements were to lead to a failed unitarity
test, then the model would be demonstrably incomplete, indicating the need for new physics.

Without the $^{38}$Ca superallowed transition, there are already thirteen $0^+$$\rightarrow$$\,$$0^+$ transitions whose
$ft$ values are known to $\sim$ 0.1\% precision \cite{Ha09}.  What is the motivation to add a fourteenth?  First, 
$^{38}$Ca decay involves states exclusively within the $sd$ shell, where model calculations are reliable, so the
isospin-symmetry-breaking correction required to extract $V_{ud}$ can be calculated with very little uncertainty
originating from nuclear-structure ambiguities.  Second, the calculated value for this correction to the $^{38}$Ca
transition is larger than it is for any of the other measured transitions with isospin-symmetry-breaking corrections of
similar reliability; if it produces a result for $V_{ud}$ that is consistent with previous values, then it serves to
confirm the veracity of the corrections for all such transitions.  Third, the $^{38}$Ca$\rightarrow$$^{38m}$K
transition is mirror to the already well-known $^{38m}$K$\rightarrow$$^{38}$Ar superallowed transition. The ratio of
mirror $ft$ values is very sensitive to the details of the isospin-symmetry-breaking calculation, and is therefore a
critical test of these calculations \cite{Pa14}. Until now, no pair of mirror $0^+$$\rightarrow$$\,$$0^+$ transitions
was known with sufficient precision to be useful in this context.

The measurement reported here is the first precise measurement of the branching ratios for the $\beta$ decay of
$^{38}$Ca.  Our result for the $0^+$$\rightarrow$$\,$$0^+$ transition and its impact on the isospin-symmetry-breaking
corrections have been reported recently in letter format \cite{Pa14}.  Here we describe the details of the experiment
itself and present information on all the $\beta$-decay branches from $^{38}$Ca.  In addition to the superallowed
branch, there are five allowed Gamow-Teller branches to $1^+$ states in $^{38}$K.  Their intensities will be seen to
agree well with $sd$-shell-model calculations, further confirming the efficacy of the model in this mass region.

Our measurement consisted of repetitive cycles, in which we deposited pure samples of $^{38}$Ca ($t_{1/2}\,$=$\,$444 ms),
moved them rapidly to a shielded counting location and recorded
$\beta$-$\gamma$ coincidences from their decay.  By measuring the absolute intensity of the $\gamma$-ray peaks in the
coincidence spectrum and comparing them to the total number of detected $\beta$ particles, we could derive the
branching ratios for the $\beta$ transitions that populated the $\gamma$-emitting states in $^{38}$K.

\section{Experiment}
\label{sec:experiment}
We produced $^{38}$Ca via the $p$($^{39}$K, 2$n$)$^{38}$Ca reaction using a 30$A$-MeV $^{39}$K primary beam from the
K500 superconducting cyclotron at Texas A\&M University.  The target was liquid-nitrogen-cooled hydrogen contained
at 2.0 atm pressure in a gas cell located in the target chamber of the Momentum Achromat Recoil Spectrometer (MARS)
\cite{Tr02}. The fully stripped reaction ejectiles were spatially separated by their charge-to-mass ratio, $q/m$, in
MARS, leaving a nearly pure $^{38}$Ca beam to emerge from the focal-plane extraction
slits.  This beam then exited the vacuum system through a 51-$\mu$m-thick Kapton window, passed through a 0.3-mm-thick
BC-404 scintillator, and then through a stack of aluminum degraders, finally stopping in the 76-$\mu$m-thick aluminized
Mylar tape of a fast tape-transport system. 

\begin{figure}[t]
\centering
\includegraphics[width=\columnwidth]{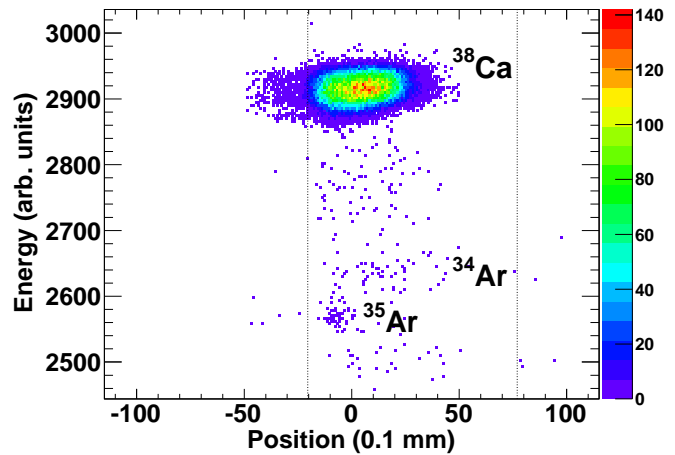}
\caption{\label{fig1} The deposited energy versus position as obtained with the PSSD in the MARS focal plane. The spectrometer
had already been optimized for $^{38}$Ca production. Vertical lines show the position of the extraction slits in MARS during
these measurements; note that the slits were set asymmetrically in order to minimize the contribution from impurities.}
\end{figure}

To optimize beam purity before the measurement began, we inserted a 1-mm-thick 16-strip position-sensitive silicon
detector (PSSD) at the MARS focal plane.  Then, working with a low-current primary beam we focused the $^{38}$Ca
beam, and also identified and minimized nearby reaction products that could reduce the purity of the beam.  The result
we obtained after this initial tuning is shown in Fig. \ref{fig1}.  With the focal-plane acceptance slits set as
indicated in the figure, the residual impurities were very weak, with the most prominent among them being $^{35}$Ar and
$^{34}$Ar.  Being of comparable mass to $^{38}$Ca, these impurities passed through the degraders and, like $^{38}$Ca, were
stopped in the tape. Lighter-mass impurities, which are not shown in the figure, have substantially longer ranges and
consequently were not stopped in the Mylar tape; so they played no role in the measurement.  Ultimately, the sample
collected in the tape was 99.7\% pure $^{38}$Ca.

During our measurement, we checked the composition of the beam daily by re-inserting the PSSD at the MARS focal plane,
and recording the spectrum of deposited energy versus position each time. No appreciable changes were observed throughout
the experiment.  All these spectra were saved and used in a detailed off-line analysis of impurities. 

Once the beam-tuning procedures had been completed, the PSSD was removed from the beam path and the primary beam intensity
was increased.  The data-taking was in repetitive cycles.  First, $^{38}$Ca was collected in the tape for 1.6 s, its
rate of accumulation being measured by the BC-404 scintillator located at the exit of MARS. Then the beam was interrupted
and the tape moved the sample in 200 ms to a shielded counting location 90 cm away, where decay data were acquired for 1.54 s. 
After counting was complete, the beam was restored and the cycle repeated.  This computer-controlled cycle,
$collect$-$move$-$count$, was repeated over 60,000 times to obtain the desired statistics.

At the counting location, a 1-mm thick BC-404 scintillator for the detection of $\beta^{+}$ particles was located 3 mm
from one side of the collection tape, and our specially calibrated high-purity germanium (HPGe) detector for $\gamma$ rays
was 151 mm away from the other side of the tape. The distance between the stopped tape and the HPGe detector was measured
during the counting period of each cycle with a laser-ranging device \cite{acuity} mounted next to the HPGe detector.  The
result, which was accurate to $\pm$5 $\mu$m, was recorded with the data for that cycle.  The measured distances were quite
consistent, with the full width at half maximum (FWHM) of their distribution being 0.5 mm.  Since our HPGe-detector efficiency
has been precisely calibrated at a source-detector distance of exactly 151.0 mm, we used the laser result to adjust the
calibrated detector efficiency to correspond with the actual average source-detector distance.  The latter was within 0.1 mm of
151.0 mm, so the adjustment was very small.

During the measurement, our data-acquisition system generated a ``master trigger'' by identifying the arrival of a $\beta$
particle and a $\gamma$ ray within $\sim$2 $\mu$s of one another.  This signaled the occurrence of a $\beta$-$\gamma$ coincident
event and initiated acquisition.  For each such event, we recorded the detected energy of both the $\beta$ and $\gamma$ rays, the exact
time difference between their arrivals, and the time that the event itself occurred relative to the beginning of the counting
period.  For each cycle we also recorded the rate of accumulation of $^{38}$Ca ions in the tape as a function of time, the
total number of $\beta$- and $\gamma$-ray singles, and the laser distance readings. The same discriminator signals used to
scale the $\beta$ singles were also used in creating the master triggers and establishing the occurrence of $\beta$-$\gamma$
coincidences. Electronic dead times for the coincidence channel and the two singles channels were measured continuously
throughout the measurement with pulser signals from a constant-frequency pulse generator being recorded in coincidence with
the gating signals from each channel.  

Room background was measured during the experiment to examine its contribution both to the $\beta$-$\gamma$ coincidence spectrum
and to the $\beta$-singles rate. We did this by using measurement cycles that were normal in every way except that the tape
motion was disabled, so that the collected sample never reached the counting location.  Under these conditions, essentially no
$\beta$-$\gamma$ coincidences were observed, and the $\beta$-singles rate dropped to 0.014\% of the rate observed under normal
conditions. Though very low, this room-background rate for $\beta$-singles was incorporated into our analysis. 

Immediately after the $^{38}$Ca branching-ratio measurement concluded, we made an off-line measurement using a 19-kBq
uncovered $^{22}$Na source placed at the tape position in the counting location.  Except that the beam was off and the tape
stationary, the configuration was identical to that used in the $^{38}$Ca measurement.  The $^{22}$Na source was chosen because
it is a positron emitter, which populates only a single $\gamma$-emitting state at 1275 keV in the daughter.  Consequently it
provides a clean view of the Compton-scattering distribution from 511-keV photons in the region around 328 keV, the energy of one of the
$\gamma$-ray peaks from $^{38}$Ca decay.  This information was invaluable to us in extracting the area of the 328-keV peak,
since that peak lies just at the edge of the Compton distribution.

\section{Analysis}
\label{s:anal}

\begin{figure}[t]
\centering
\includegraphics[width=0.75\columnwidth]{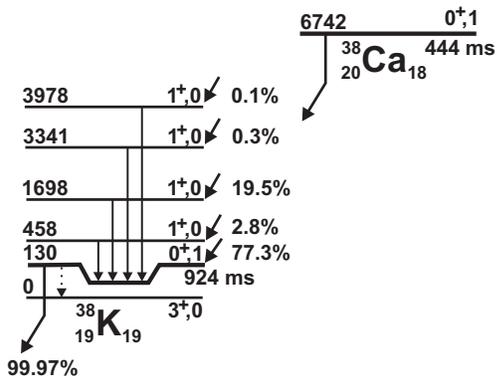}
\caption{\label{fig2} Beta-decay scheme of $^{38}$Ca showing the four strongest $\beta$ branches and the most intense subsequent $\gamma$-ray
transitions in $^{38}$K. For each level, its (J$^{\pi}$, T) is given as well as its energy, expressed in keV, relative to the $^{38}$K ground
state \cite{Ca08}.  Branching percentages come from this measurement.
As discussed later in Sec.~\ref{ss:relgamma}, there are other weak $\gamma$ rays present, which complicate this scheme somewhat.  Note that
there is also a very weak (0.03\%) $\gamma$-ray branch from the 130-keV isomer to the ground state \cite{Le08}.}
\end{figure}

A simplified $\beta$-decay scheme for $^{38}$Ca is shown in Fig.~\ref{fig2}, from which it can be seen that the superallowed transition
directly feeds the 130-keV isomer in $^{38}$K.  This state has a half-life of 924 ms, so no prompt $\gamma$ ray is emitted following
its population.  In contrast, all the allowed Gamow-Teller transitions to higher excited $1^+$ states are followed by prompt
$\gamma$ rays, predominantly emitted in each case directly to the isomeric state. It is these latter transitions whose absolute
intensity we can measure from the $\beta$-coincident $\gamma$-ray spectrum.  The fact that the observed transitions only account
for a relatively small percentage, $\sim$23\%, of the total decay, actually works to our advantage.  Since we obtain the superallowed
branching ratio by subtracting the total of the Gamow-Teller branching ratios from 100\%, the relative uncertainty of
the result is reduced by a multiplicative factor of 0.3 (=23/77) compared to that of the sum of the measured Gamow-Teller branches.
A measurement precision of $\pm$0.6\% for the Gamow-Teller branches leads to a $\pm$0.2\% result for the superallowed branch. 

To determine the branching ratio for the superallowed transition from $^{38}$Ca precisely, our first step was to establish the
$\beta$-branching ratio to the 1$^{+}$ state in $^{38}$K at 1698 keV, the most intense branch observed. We accomplished this by
obtaining the number of $\beta$-coincident 1567-keV $\gamma$ rays relative to the total number of positrons emitted
from $^{38}$Ca. Then, combining the relative intensities of all the other (weaker) observed $\gamma$-ray peaks, we determined the
total Gamow-Teller $\beta$-branching to all 1$^{+}$ states. Finally, the subtraction of this total from 100\% resulted in the
branching ratio for the superallowed transition.

To be more specific about our first step, we write the $\beta^+$-branching ratio, $R_i$, for a pure $\beta^+$-transition populating the
particular state $i$, which is de-excited by the emission of a single $\gamma$ ray, $\gamma_i$, as follows:
\begin{equation}
\label{eq:branchingratio}
R_i = \frac{N_{\beta\gamma_i}}{N_{\beta}\ \epsilon_{\gamma_i}} \frac{\epsilon_\beta}{\epsilon_{\beta_i}}~,
\end{equation}
where $N_{\beta\gamma_i}$ is the total number of $\beta$-$\gamma$ coincidences in the $\gamma_i$ peak; $N_{\beta}$ is the total
number of beta singles corresponding to $^{38}$Ca $\beta$ decay; $\epsilon_{\gamma_i}$ is the efficiency of the HPGe detector for
detecting $\gamma_i$ rays; $\epsilon_{\beta_i}$ is the efficiency of the plastic scintillator for detecting the betas that populate
state $i$; and $\epsilon_\beta$ is the average efficiency for detecting the betas from all $^{38}$Ca transitions.  Note that this
equation applies only to positron emission.  Although the contribution from electron capture for $A$=38 is very small, it must
be accounted for.  Furthermore, small corrections must be applied to incorporate the effects of observed weak $\gamma$
transitions between the states in $^{38}$K.  Both these adjustments will be dealt with in Sec.~\ref{s:results}.

In the immediately following sections, after describing some initial processing of the experimental data, we detail how all the factors on the
right-hand side of Eq.~(\ref{eq:branchingratio}) were obtained specifically for the $\beta$ transition to the 1698-keV state. 

\subsection{Cycle selection}
\label{ss:cyc}

Before analysis began, we filtered our accumulated data by rejecting cycles that did not meet certain criteria. The first criterion
we applied was the number of implanted $^{38}$Ca ions detected by the BC404 scintillator at the exit of MARS during each collection
period. We accepted only those cycles in which the collection rate lay between 4200 and 24,000\footnote{In Ref.~\cite{Pa14}, this upper
limit was incorrectly stated to be the average implantation rate of $^{38}$Ca ions into the tape.  In fact, the average rate was
15,000 ions/s for the cycles used in our final analysis.} ions/s.  This removed cycles that had very little -- or no -- primary
beam from the cyclotron, as well as those with abnormally high beam current, which could have adversely affected our system dead time.

Our second criterion was based on the ratio of the number of $\beta$-particles detected to the number of $^{38}$Ca ions implanted for
each cycle. We set limits on this ratio to reject cycles in which the tape-transport system had not moved the sample into the designated
counting position between the $\beta$ detector and the HPGe detector, thus significantly reducing the counting rate for $\beta$
particles (though not appreciably reducing the singles $\gamma$-ray rate).  A third criterion could have been applied: the reading from
our laser measurement of the distance between the tape and the HPGe detector.  However, we found that once the second criterion had been
applied, these distance measurements were symmetrically distributed within a narrow range (see Sec.~\ref{sec:experiment}), so no
additional filtering was necessary.

In the end, our selection criteria provided 60,847 good cycles, $\sim$89\% of the total cycles recorded.  These good cycles contained
approximately 8.8 million $\beta$-$\gamma$ coincidences, corresponding to over $3.7 \times 10^8$ $\beta$ singles.  All subsequent analysis
incorporates only the data from these cycles. 

\begin{figure}[t]
\centering
\includegraphics[width=0.8\columnwidth]{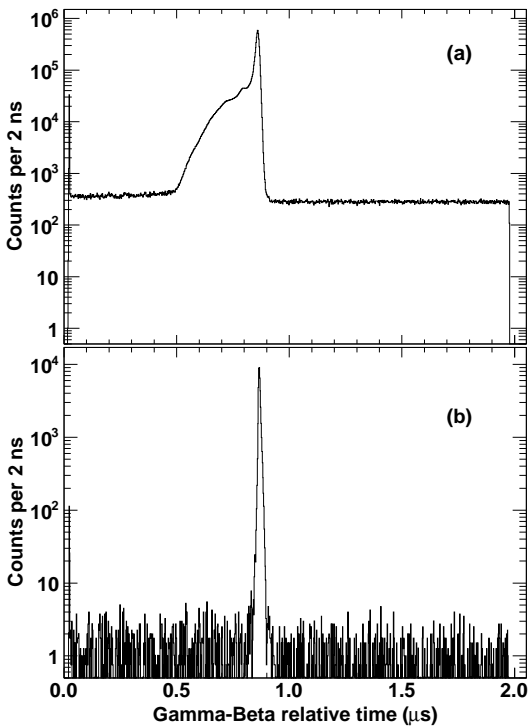}
\caption{\label{fig3} (a) Spectrum of measured time-differences between the arrival of a $\gamma$ ray and that of a
$\beta$ particle for all identified coincidence events.  Note that the $\beta$ signal was electronically delayed so that the
prompt-coincidence peak appears near the center of the spectrum.  (b) Measured time-difference spectrum for events corresponding only
to the 1567-keV $\gamma$ ray.}
\end{figure} 

\subsection{Eliminating random coincidences}
\label{ss:rand}

For each event, we recorded the time between the detection of a $\gamma$-ray and the subsequent arrival of an electronically delayed signal
from the positron detector, as measured with a time-to-digital
converter (TDC).  This time spectrum for all events identified by our master trigger appears in the top panel of Fig.~\ref{fig3}.  The broad peak
corresponds to real ($i.e.$ ``prompt'') coincident events, while the flat distributions to the left and right are from random coincidences. 
The single-channel peak at zero is an artifact resulting from the way in which we establish the master trigger: It contains only random coincidences, their
number being proportional to the time-width of the $\beta$ signal we used to establish the existence of a coincidence.

\begin{figure}[b]
\centering
\includegraphics[width=\columnwidth]{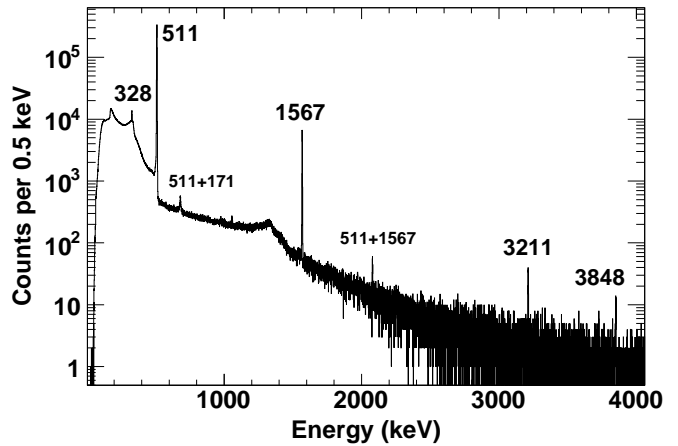}
\caption{\label{fig4} Spectrum of $\gamma$ rays observed in prompt coincidence with positrons from the decay of $^{38}$Ca. The small peak
labeled ``511+171" is the sum of two annihilation photons, one of which has backscattered into the detector.  The ``511+1567" peak is the
result of coincidence summing between a 1567-keV $\gamma$ ray and annihilation radiation from the positron decay that preceded it.}
\end{figure}

An interesting curiosity, not often seen in coincidence time spectra, is the difference in the random-coincidence rates before and after the
prompt peak.  This is clearly visible in our measurement because of the high, $\sim$40\%, efficiency of our $\beta$ detector.  Since this
percentage of the total $\gamma$-ray signals were removed by the occurrence of a prompt coincidence, the rate after the prompt peak is only
60\% of the rate before.

Based on this time spectrum, we could easily produce a $\gamma$-ray spectrum free of random-coincidence events.  We did this by first gating
on the part of the time spectrum that contains the prompt peak and then gating on the flat, random parts of the spectrum.  The $\gamma$-ray
spectrum obtained from the second gate, suitably normalized, was then subtracted from the spectrum from the first gate.  The resultant
spectrum appears in Fig.~\ref{fig4}.  It exhibits four clear $\gamma$-ray peaks, at 328 keV, 1567 keV, 3211 keV, and 3848 keV, as well as
several coincidence summing peaks, all of which are related to the decay of $^{38}$Ca.  No other peaks are immediately visible.

The prompt peak appearing in the top panel of Fig.~\ref{fig3} has a noticeable tail to the left.  This occurs because it
includes all coincident events, which means the full range of detected $\gamma$-ray energies is represented (see Fig.~\ref{fig4}).  Low-energy
$\gamma$ rays trigger the TDC later than higher energy ones, and a later $\gamma$-ray trigger results in a shorter time before the corresponding
$\beta$-particle arrives: hence the tail to shorter times.  The bottom panel of Fig.~\ref{fig3} shows the time spectrum corresponding to a
single $\gamma$-ray peak, the one at 1567 keV.  The prompt peak in this case is much narrower (FWHM $<$10 ns) and has no tail.  It is spectra
like this, each restricted to a narrow energy window around a single $\gamma$-ray peak, that we used in our final analyses for the contents of
individual peaks.

\subsection{Efficiency calibrations}
\label{ss:eff}

From the appearance of $\epsilon_{\gamma_i}$ and the ratio $\epsilon_\beta/\epsilon_{\beta_i}$ in Eq.~(\ref{eq:branchingratio}) it is
clear that our determination of the superallowed branching ratio must rely on a precise absolute efficiency calibration of the
$\gamma$-ray detector, and a reasonable knowledge of relative efficiencies for the $\beta$ detector.

Our HPGe detector has been meticulously calibrated at a source-to-detector distance of 151 mm.  This was reported thoroughly a decade ago
\cite{He03, He04}.  Originally we covered the energy range from 22 to 1836 keV, taking data from 13 individual sources of 10 different
radionuclides: $^{48}$Cr, $^{60}$Co, $^{88}$Y, $^{108m}$Ag, $^{109}$Cd, $^{120m}$Sb, $^{133}$Ba, $^{134}$Cs, $^{137}$Cs and $^{180m}$Hf
\cite{He03}.  Of crucial importance were two $^{60}$Co sources specially prepared by the Physikalisch-Technische Bundesanstalt \cite{Sc02},
with activities certified to $\pm$0.06\%.  These sources were used to anchor our absolute efficiency calibration, with cascaded $\gamma$-ray transitions
from the other sources providing precise links over a wide range of energies.  At the time, in addition to acquiring calibration spectra, we
also made a number of measurements designed to reveal the physical dimensions and location of the detector's Ge crystal in its housing.  This
information was then used as input to Monte Carlo calculations performed with the electron and photon transport code CYLTRAN \cite{Ha92}.
With only the thicknesses of the detector's two dead layers as adjustable parameters we achieved excellent agreement ($\chi^2/N = 0.8$) between the Monte Carlo
efficiency results and our 40 measured data points.  A year later, using $^{24}$Na, $^{56}$Co and $^{66}$Ga sources we extended our region
of calibration up to 3.5 MeV \cite{He04}.

Ever since these calibrations were made, we have kept the detector continuously at liquid-nitrogen temperature to ensure that the internal dead
layer did not expand.  We have also periodically re-measured one of the precisely calibrated $^{60}$Co sources.  Some slight tailing has
appeared on the $^{60}$Co peaks but, if these tails are included in the peak area, no change in detector efficiency can be detected.  As a result,
we can continue to use CYLTRAN calculations to obtain our detector efficiency with $\pm$0.2\% uncertainty in the range 50-1800 keV, and with
$\pm$0.4\% from 1800 to 3500 keV.  The efficiencies obtained for the four main $\gamma$-rays of interest are listed in the third column of
Table~\ref{t:tab1}.  Note that the weakest $\gamma$ ray of the four, at 3848 keV, lies slightly above the range of calibration,
so its efficiency carries a somewhat larger uncertainty.  The peak is so weak, however, that its efficiency uncertainty is completely swamped
by its statistical uncertainty.

\begin{table}
\caption{\label{table1} Detector efficiencies are given for the $\gamma$ rays, $\gamma_i$, that de-excite states $E_{x_i}$ in $^{38}$K; and for $\beta$ particles
emitted in the decay branches, $\beta_i$, that populate states $E_{x_i}$.  The values for $\epsilon_{\gamma_i}$ apply to our HPGe detector, and the ratios,
$\epsilon_\beta/\epsilon_{\beta_i}$, to our thin $\beta$ scintillator.
\label{t:tab1}}
\begin{ruledtabular}
\begin{tabular}{lllll}
& & & & \\[-3mm]
\multicolumn{1}{l}{$E_{x_i}$\footnotemark[1]} & \multicolumn{1}{l}{$E_{\gamma_i}$\footnotemark[1] for} & & $E_{\beta max}$ for & \\
\multicolumn{1}{l}{in $^{38}$K}
 & \multicolumn{1}{l}{$\gamma$ decay}
 & \multicolumn{1}{c}{$\epsilon_{\gamma_i}$}
 & \multicolumn{1}{l}{$\beta_i$ feeding}
 & \multicolumn{1}{l}{$\epsilon_\beta/\epsilon_{\beta_i}$} \\
\multicolumn{1}{l}{(keV)}
 & \multicolumn{1}{l}{(keV)}
 & \multicolumn{1}{c}{(\%)}
 & \multicolumn{1}{l}{(keV)}
 & 
 \\[1mm]
\hline
& & & & \\[-3mm]
   130.4 & ~~~~--                                    
   & ~~~~-- & 5590.1  &   0.9989  \\
   458.5 &                                     
   327.9  &  $0.5401(11)$  &
   5262.0  &   0.9988  \\
   1697.8 &                                     
   1567.4  &  $0.1777(4)$  &
   4022.7  &   1.0038  \\
   3341.2  &                                    
   3210.7  &  $0.0974(4)$  &
   2379.3  &   1.0267  \\
   3977.3  &                                   
   3848.0  &  $0.0816(6)$  &
   1743.2  &   1.0860  \\
\end{tabular}
\end{ruledtabular}
\footnotetext[1]{Values taken from Ref.~\cite{Ca08}.}
\end{table}

Our $\beta$ detector consists of a 1-mm-thick Bicron BC404 scintillator disc recessed into a cylindrical Lucite light guide, which is coupled
in turn to a photomultiplier tube.  The response function of this detector has been extensively characterized as a function of $\beta$-particle
energy by a combination of GEANT4 Monte Carlo simulations \cite{Ag03} and measurements with $^{133}$Ba, $^{137}$Cs and $^{207}$Bi sources, all three
of which emit both $\beta$-decay electrons and conversion electrons.  The agreement between measurements and simulations was found to be excellent \cite{Go08}.
Since those studies were completed six years ago, we have extended our source tests to $^{22}$Na, with similar success; $^{22}$Na, like $^{38}$Ca,
is a positron emitter.  We have also demonstrated that the EGSnrc Monte Carlo code \cite{EGS} produces equally good agreement with measurements
and runs more rapidly than GEANT4, so we have used the former code in the present analysis.

As is clear from Eq.~(\ref{eq:branchingratio}), we do not need to know the absolute efficiency of our $\beta$ detector but rather how the efficiency
changes as a function of the end-point energy, $E_{\beta max}$, which is different for each $\beta$-decay branch that feeds a state in
$^{38}$K.  The energy dependence of our $\beta$-detection efficiency is caused principally by the fixed low-energy electronic threshold,
which removes a slightly different fraction of the total $\beta$ spectrum for different end-point energies.  This is important because any difference
in the $\beta$-detection efficiency from one $\beta$ transition to another affects the measured intensity of coincident $\gamma$ rays following
that $\beta$ transition.

\begin{figure}[t]
\centering
\includegraphics[width=\columnwidth]{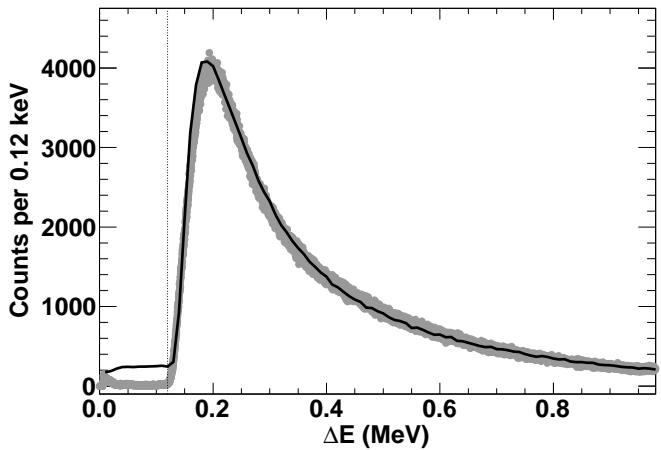}
\caption{\label{fig5} The measured energy deposition (filled circles, gray) in the $\beta$-detector for the decay of $^{38}$Ca is compared with the
EGSnrc-simulated spectrum (solid line, black). The dashed vertical line at 120 keV indicates our threshold.}
\end{figure}

Figure \ref{fig5} presents the $\beta$-detector energy spectrum as measured for $^{38}$Ca decay, along with a Monte-Carlo spectrum generated with
EGSnrc.  We have included the transport tape, together with the rest of the nearby counting-location geometry in the simulation.  Clearly, the shape
of the simulated spectrum is in a good agreement with the measured one, thus giving us confidence that we can obtain reliable efficiency ratios,
$\epsilon_\beta/\epsilon_{\beta_i}$ for the transitions of interest.  The results for the strongest transitions are shown in the fifth column of
Table~\ref{t:tab1}.  They appear without uncertainties since the results are all quite near unity so the statistical uncertainties due to the
Monte Carlo calculations are negligible in the present context.  The calculated absolute efficiency for the total decay of $^{38}$Ca,
$\epsilon_{\beta}$, is approximately 40\%.  Its precise value is not required.

\subsection{Beta singles}
\label{ss:bsingles}

The $N_{\beta}$ term in Eq.~(\ref{eq:branchingratio}) of course refers only to the $\beta$ particles emitted from $^{38}$Ca.  However, the number we
obtain from our $\beta$ detector includes not only the $\beta$'s from $^{38}$Ca but also those from its daughter $^{38}$K.  In addition there are other
much smaller contributions, including those from any weak impurities that remain in the collected samples.  We deal first with the impurities.

\subsubsection{Impurities}
\label{sss:imp}

Based on the spectrum of reaction products detected at the MARS focal plane
(see Sec.~\ref{sec:experiment}), we determined the initial intensity of weak contaminants; then we used the SRIM code \cite{Zi08} to calculate the
energy loss each would experience in passing through the degraders, and thus derived the amount of each that would be collected in our tape.  Expressed
as a percentage of the $^{38}$Ca deposited, we found the initial intensities in the tape to be: $^{37}$K -- 0.02(1)\%; $^{36}$K -- 0.03(2)\%;
$^{35}$Ar -- 0.20(10)\%; $^{34}$Ar -- 0.07(2)\%; and $^{34}$Cl -- 0.03(1)\%.  Naturally, each of these activities has a different half-life, and some
populate $\beta$-decaying daughters as well. All this information is well known, though, so we have used it to calculate that the total of all
impurities contributed 0.58(30)\% to the total number of $\beta$'s recorded during the counting period.

Most of these impurities have only very weak branches to $\gamma$-emitting states so, for them, there is no evidence available to corroborate
these calculations.  The decay of $^{36}$K is the sole exception: Its $\beta$ decay leads to the emission of a 1970-keV $\gamma$ ray 82\% of the time
\cite{Ni12}.  This $\gamma$-ray peak is weakly present in our $\beta$-coincident $\gamma$-ray spectrum, allowing us to determine the ratio of its area
to that of the 1567-keV $\gamma$-ray peak from $^{38}$Ca decay.  The result, though imprecise, is consistent with the relative intensity obtained from
the MARS focal plane spectrum. 

\subsubsection{Parent $\beta$-decay fraction}
\label{sss:pdimp}

Much more significant is the contribution to the measured $\beta$ singles from the decay of $^{38m}$K, the daughter of $^{38}$Ca. This nuclide is not present in the
implanted beam, but it naturally grows in the collected sample as the $^{38}$Ca decays. Since the half-lives of $^{38}$Ca and $^{38m}$K are well known
\cite{Bl10,Pa11,Ha09}, the ratio of their activities can easily be calculated if the $^{38}$Ca implantation rate is known as a function of time during
the collection period \cite{Pa11}.  To enable this calculation, we recorded the time profile of ions detected by the scintillator at the exit of MARS
for each individual cycle.  A typical result for a single run, comprising 2,277 cycles, appears as Fig.~\ref{fig6}.  Before making the calculation,
though, we need to correct, if necessary, for three more small effects:

\begin{figure}[b]
\centering
\includegraphics[width=0.99\columnwidth]{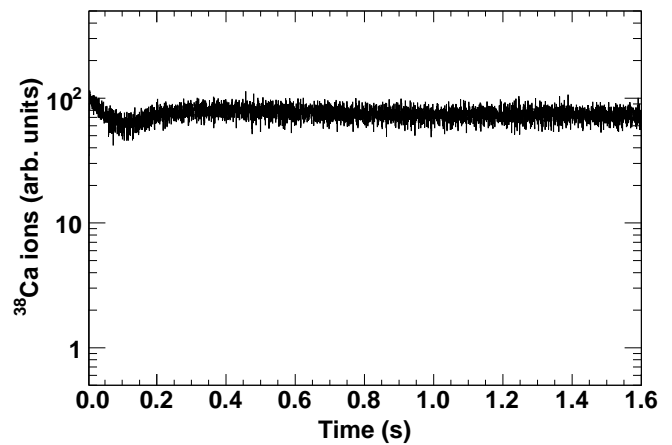}
\caption{\label{fig6}Typical time-profile of the collected $^{38}$Ca beam measured over the course of one run. 
The initial drop in intensity is generated by the decrease in local density of the hydrogen in the target cell
as the primary beam heats the gas around its path.  A fan located inside the gas-target mitigates the effect and
ensures a rapid transition to stable conditions.}
\end{figure}

\begin{enumerate}[(i)]

\item  {There is a tiny probability that $\gamma$ rays produced in the decay of $^{38}$Ca get counted in the $\beta$ scintillator.  This does not matter for
        annihilation radiation, which can be thought of as a surrogate for a $\beta$ particle, since it would not alter the efficacy of Eq.~(\ref{eq:branchingratio}).
        However it does matter for the transition $\gamma$ rays in cases where they are detected but the corresponding $\beta$ particle that feeds the transition is
        not.  Using EGSnrc Monte Carlo simulations, we determined that 0.043(4)\% of the total counts in the $\beta$ detector are $\gamma$ rays of this type.  We
        reduced the recorded number of counts in the detector by this factor.}

\item  {The decay of $^{38m}$K includes a weak 0.033(4)\% $\gamma$-ray branch to the ground state of $^{38}$K \cite{Ha09}, so the number of observed $\beta$'s
        must be less than the total decays of $^{38m}$K by that amount.  We apply this correction.}

\item  {Although the $E_{\beta max}$ for $^{38m}$K decay is less than it is for the superallowed transition from $^{38}$Ca, the other lower-energy 
        branches from $^{38}$Ca act to offset the efficiency difference that would otherwise be expected (see discussion in Sec.~\ref{ss:eff}).  As a result,
        the total efficiency for observing $\beta$'s from $^{38m}$K is very nearly the same as it is for observing them from $^{38}$Ca.  Nevertheless, the
	remaining 0.025\% efficiency difference was incorporated.}
 
\end{enumerate}

We are now in a position to calculate what fraction of the true $A$=38 $\beta$-decay events recorded in our detector is due to $^{38}$Ca decay.  Using the
time profile of the $^{38}$Ca collection rate (see Fig.~\ref{fig6}), together with the known half-lives of $^{38}$Ca and its daughter $^{38m}$K, we calculated
the total number of decays of each, integrated over the precisely delineated counting period.  The fraction attributable to $^{38}$Ca was thus found to be 0.3510(3).
This result incorporates the corrections described in items (ii) and (iii) above.

\subsubsection{Final result for $N_{\beta}$} 

To obtain our final result for $N_{\beta}$ from the total counts recorded in our $\beta$ detector, we first remove room background, and then correct for $\beta$-decay
events from impurities and for $\gamma$ rays counted in the $\beta$ detector.  Finally we apply the calculated $^{38}$Ca $\beta$ fraction.  These steps appear
quantitatively in Table~\ref{table2}, together with the final result for $N_{\beta}$. 

\begin{table}[t]
\caption{\label{table2}Derivation of $N_{\beta}$ from the total number of singles events recorded in
the $\beta$ detector }
\begin{ruledtabular}
\begin{tabular}{lll}
Quantity   &  Value  & Source  \\
\hline \\[-2mm]
Total $\beta$-detector counts & $3.72623(19)\times 10^8$ &  \\
Background  & $-5.060(23)\times 10^4$  & Sec.~\ref{sec:experiment}  \\
$\beta$-decay of impurities  & $-0.58(30)\%$  & Sec.~\ref{sss:imp} \\
Detected $\gamma$ rays  &  $-0.043(4)\%$  &  Sec.~\ref{sss:pdimp} \\
$^{38}$Ca fraction of $\beta$'s \footnotemark[1] & $\times0.3510(3)$  & Sec.~\ref{sss:pdimp} \\
\cline{2-2} 
& &   \\[-2mm]
$N_{\beta}$($^{38}$Ca)  & $1.2996(40)\times 10^8$ &  \\
\vspace{-10.pt}
\footnotetext[1]{Calculation takes account of the weak (0.033\%) $\gamma$-decay branch from $^{38m}$K. }
\end{tabular}
\end{ruledtabular}
\end{table}

\subsection{$\beta$-coincident 1567-keV $\gamma$ rays}
\label{ss:Ngamma}

The number of $\beta$-coincident 1567-keV gamma rays, $N_{\beta\gamma_{1567}}$, is primarily determined from the integrated area of the 1567-keV
$\gamma$-ray peak recorded in coincidence with the prompt peak in the $\gamma$-$\beta$ time spectrum: See Fig.~\ref{fig3} and the description
in Sec.~\ref{ss:rand}.  Our procedure for extracting all peak areas was to use a modified version of GF3, the least-squares peak-fitting program in
the RADWARE series \cite{Ra}. A Gaussian peak with a smoothed step function and a linear background in the peak region were sufficient to properly
describe the detailed shape of all peaks of interest in the spectrum. This was the same fitting procedure as was used in the original
detector-efficiency calibration \cite{He03, He04}.  The result obtained for the 1567-keV peak is given in the top row of Table~\ref{table3}.

Before this result can be used in Eq.~(\ref{eq:branchingratio}), there are several small corrections that must be applied to account for coincidence
summing, dead-time and pile-up.  These corrections are described in the following sections.

\subsubsection{Coincidence summing}
\label{sss:coinc}

The 1567-keV $\gamma$ rays are emitted from a state at 1698 keV in $^{38}$K, following a positron decay branch from $^{38}$Ca that populates the
state (see Fig.~\ref{fig2}). To be recorded in our measurement, the positron must have appeared in our $\beta$ detector in coincidence with the
$\gamma$ ray in our HPGe detector.  Since the positron generally annihilates in or near the $\beta$ detector, there is a non-negligible probability
that one of the resulting 511-keV photons will also be recorded in the HPGe detector, and will sum with the 1567-keV $\gamma$ ray, thus removing
some of the latter events from the full-energy peak.  The resultant sum peak at 2078 keV (1567 + 511) is clearly visible, though small, in our
$\beta$-coincident $\gamma$-ray spectrum in Fig.~\ref{fig4}.  

Unfortunately the sum peak at 2078 keV only accounts for a portion of the total events lost from the peak at 1567 keV.  To determine the total losses,
we must incorporate the complete 511-keV response function since losses from the 1567-keV peak also result from summing with signals from 511-keV photons
that Compton scatter in the HPGe crystal and deposit less than their full energy.  What we need to know then is the ratio of the total efficiency of
our detector to its full-energy-peak efficiency -- the total-to-peak ratio -- for 511-keV photons.

In the original calibration of our detector, we measured the total-to-peak ratio as a function of energy \cite{He03}; but this was in an open geometry
without the shielding necessitated by an in-beam measurement.  In principle, the full response function under measurement conditions could be extracted
from the first 511 keV of the spectrum in Fig. \ref{fig3} but, in practice, the spectrum is distorted there by the response to the 328-keV and 1567-keV
$\gamma$ rays.    

\begin{table}[t]
\caption{\label{table3}Derivation of $N_{\beta\gamma_{1567}}$ from the total number of events in the 1567-keV peak in the $\beta$-coincident $\gamma$-ray spectrum. }
\begin{ruledtabular}
\begin{tabular}{lll}
Quantity   &  Value  & Source  \\
\hline \\[-2mm]
Area of 1567-keV peak & 42,941(211) &  \\
511-keV summing   & +1,121(91)  & Sec.~\ref{sss:coinc}  \\
bremsstrahlung summing  & +73(22)  & Sec.~\ref{sss:coinc} \\
Dead time/pile-up  &  $\times1.0093(7)$  &  Sec.~\ref{sss:dead} \\
Random preemption & $\times1.0023(4)$  & Sec.~\ref{sss:preempt} \\
\cline{2-2} 
& &   \\[-2mm]
$N_{\beta\gamma_{1567}}$  & 44,648(242) &  \\
\vspace{-10.pt}
\end{tabular}
\end{ruledtabular}
\end{table}

We therefore established the required response function by using an off-line source of $^{68}$Ge.  This nuclide decays by electron capture to $^{68}$Ga,
which in turn decays by positron emission (and electron capture) almost entirely to the ground state of $^{68}$Zn.  The resulting HPGe spectrum is thus dominated
by annihilation radiation with only a 1.6\% relative contribution from a 1077-keV $\gamma$-ray transition.  To reproduce experimental conditions as closely as
possible, we recorded this spectrum in coincidence with detected positrons just as we had for the $^{38}$Ca measurement.  For completeness, we obtained
three separate spectra, one in open geometry, one with partial shielding, and one with all the shielding used in the on-line experimental configuration. 
As expected, as the amount of shielding increased, the total-to-peak ratio for the 511-keV radiation increased too.  It must be noted that, because the
coincident positrons are traveling away from the HPGe detector when they annihilate, this total-to-peak result also includes a contribution from
annihilation in flight.  Therefore the total-to-peak ratio obtained for $^{68}$Ga had to be adjusted upward slightly to account for the increased effects
of annihilation in flight for $^{38}$Ca decay, which has a substantially higher $\beta^+$ end-point energy.  The final total-to-peak ratio for 511-keV
radiation as it applies to $^{38}$Ca decay was determined to be 3.64(3).  

This total-to-peak ratio multiplied by the area of the 2078-keV sum peak, determined to be 308(25) counts, establishes the total losses from the
1567-keV peak due to summing with annihilation radiation to be 1121(91) counts.  This result appears as a correction to the 1567-keV peak area in
Table~\ref{table3}.

External bremsstrahlung, emitted when positrons from $^{38}$Ca stop in the $\beta$ detector or its surroundings, is another source of coincidence summing.
In contrast to the summing just described, bremsstrahlung summing does not produce a tell-tail sum peak in the $\gamma$-ray spectrum but results rather in
a continuous energy spectrum indistinguishable from the summed Compton distributions resulting from detected $\gamma$ rays. To arrive at the total contribution
from bremsstrahlung in the spectrum of Fig.~\ref{fig4}, we first took the areas of all $\gamma$-ray peaks (including the 511-keV peak), multiplied each by its
corresponding total-to-peak ratio, summed the results and subtracted the sum from the total number of counts in the spectrum.  We took this difference to be
the contribution from bremsstrahlung.  Knowing this number and the full-energy-peak efficiency of our detector for 1567-keV $\gamma$ rays, we could calculate
the probability for coincidence summing between those $\gamma$ rays and the bremsstrahlung.  We determined the resultant loss from the 1567-keV peak to be
73(22) counts or 0.17(5)\% of the total.  This amount appears as an applied correction in Table~\ref{table3}.

\subsubsection{Dead time and pile-up}
\label{sss:dead}

Dead time in the $\beta$-detection system is small -- 450 ns per event -- and it affects equally both the numerator and denominator in Eq.~(\ref{eq:branchingratio}), so it need
not be considered any further.  However, dead time and pile-up do affect the much slower signals from the HPGe detector, and their impact depends not only
on the rate of coincident $\gamma$ rays, which averaged 94 counts/s, but also on the singles $\gamma$ rate, which averaged 430 counts/s. Furthermore, the
rate during each cycle naturally decreased with time.

The dead time per event for encoded coincident $\gamma$ rays was measured on-line to be 25.6 $\mu$s; this value also encompasses the pile-up time. For
singles $\gamma$ rays, which are not encoded, the pile-up time was determined from the signal pulse shape to be 17 $\mu$s; this value subsumes the dead
time for such events.  Since both dead time and pile-up remove legitimate signals, we treat them together.  We calculated the total losses from both sources
by integrating over the whole counting period incorporating the decrease in rate caused by the decay of $^{38}$Ca, and the growth-and-decay of $^{38m}$K.  Our
result is that losses due to the combination of dead time and pile-up amount to 0.93(7)\%.  We list the required correction factor, 1.0093(7), in
Table~\ref{table3}.

\subsubsection{Random preemption of real coincidences}
\label{sss:preempt}

There is a small probability that coincidences get lost as a result of a random coincidence preempting a real one.  This can occur if a master trigger is
generated by a real $\beta$-$\gamma$ coincidence, which starts our timing clock (the ADC), but a random $\beta$ event stops the clock before the true coincident
$\beta$ does.  This effect can easily be calculated from the known rate of $\beta$ signals and the time between the clock start and the appearance of the prompt
peak: see Fig.~\ref{fig3}(b). We calculated the correction factor required to compensate for this effect to be 1.0023(4); this factor also appears in
Table~\ref{table3}.

\subsubsection{Final result for $N_{\beta\gamma_{1567}}$}

All the corrections required to obtain the value of $N_{\beta\gamma_{1567}}$ from the measured area of the 1567-keV peak are collected in Table~\ref{table3}.  We
first add back the counts lost to coincidence summing, then multiply by the correction factors accounting for dead time, pile-up and random preemption of
true coincidence events.  The resultant value for $N_{\beta\gamma_{1567}}$ appears in the last row of Table~\ref{table3}.  It is the last piece of input data
that is required to complete the right hand side of Eq.~(\ref{eq:branchingratio}).

However, Eq.~(\ref{eq:branchingratio}) in fact yields the $\beta$ branching ratio for the production of 1567-keV $\gamma$ rays.  This is only equal to the
$\beta$ branching ratio to the 1698-keV level if that state is solely populated by $\beta$ decay and de-populated by a single $\gamma$ transition.  This is
almost, but not exactly true for the 1698-keV state in $^{38}$K.  As we will find in the next section, there are other weak $\gamma$ transitions that must
be accounted for before the true $\beta$ branching ratio to this state or any of the other $1^+$ states can be established.  

\begin{table}[t]
\caption{\label{intensity}Relative intensities of $\beta$-delayed $\gamma$ rays from the $\beta^+$ decay of $^{38}$Ca.}
\begin{ruledtabular}
\begin{tabular}{clll}
& 
\multicolumn{3}{c}{$I_{\gamma}$}  \\
\cline{2-4} \\[-2mm]
\textnormal{$E_{\gamma}$ [keV]} & 
\multicolumn{1}{c}{Ref. \cite{Wi80}} & 
\multicolumn{1}{c}{Ref. \cite{An96}} & 
\multicolumn{1}{c}{This work}   \\ 
\hline \\[-2mm]
328  & 0.126(16)   & 0.150(10)   & 0.1489(26)    \\
1240 & $<$ 0.010   & 0.0024(5)   & 0.0036(13)   \\
1567 & 1                & 1               & 1     \\
1643 & $<$ 0.010   & 0.0040(5)   & 0.0010(7)  \\
1698 & $<$ 0.0082 & 0.0008(4)   & $<$ 0.0008  \\
2883 & $<$ 0.0033 & 0.007(2)     & 0.0006(4) \\
3211 & 0.0139(15) & 0.0138(10)  & 0.0150(9)   \\
3519 & $<$ 0.0042 & 0.0004(3)   & $<$ 0.0003   \\
3716 & $<$ 0.0045 & 0.0002(1)   & $<$ 0.0005  \\
3726 & $<$ 0.0036 & 0.0019(2)   & 0.0007(3)  \\
3848 & $<$ 0.0081 & 0.0056(5)   & 0.0051(7)  \\
\end{tabular}
\end{ruledtabular}
\end{table}

\subsection{Relative $\gamma$-ray intensities}
\label{ss:relgamma}

So far, in Figs.~\ref{fig2} and \ref{fig4}, we have identified only the four most prominent $\gamma$-ray peaks in the decay of $^{38}$Ca, but in fact
there are a number of other weaker ones.  To improve our statistical accuracy in characterizing these peaks, we combined the $\gamma$-ray spectrum in
Fig.~\ref{fig4} with data from an earlier $^{38}$Ca measurement of ours.  As it turned out, that measurement could not be used to determine absolute
branching ratios because we found that we had incomplete control over all the necessary parameters, but the $\beta$-coincident $\gamma$-ray spectrum is
perfectly valid for the determination of relative $\gamma$-ray intensities.  The energy calibrations of both spectra were based on their five strongest
peaks, at 328, 511, 1567, 3211, and 3848 keV; and these peaks were also used to effect the small gain adjustment that was required before the spectra
could be combined. 

Our search for, and identification of weak $\gamma$ rays following the decay of $^{38}$Ca was based on the known level scheme of $^{38}$K \cite{Ca08}
and was guided by the results of previous studies \cite{Wi80,An96}.  In the fit of the 328-keV $\gamma$-peak area, it was especially important to have a
reliable ``background'' in that energy region because the 328-keV $\gamma$-ray full-energy peak is located close to the Compton edge of the scattering
distribution from 511-keV positron-annihilation radiation. This background was obtained from the $\beta$-delayed $\gamma$-ray spectrum of a $^{22}$Na
source measured under identical conditions, as described in Sec.~\ref{sec:experiment}. For the energy range between 200 keV and 400 keV, we compared the
$\gamma$-ray spectrum of $^{22}$Na with that of $^{38}$Ca. The result verified that no unexpected structure lay beneath the 328-keV peak in the
$^{38}$Ca $\gamma$-ray spectrum.

\begin{figure}[b]
\centering
\includegraphics[width=0.9\columnwidth]{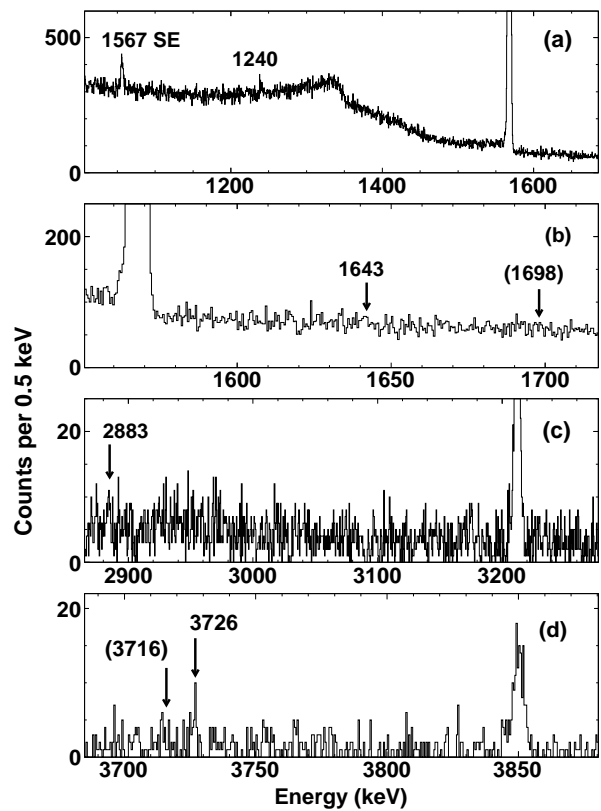}
\caption{\label{fig7}Portions of the summed $\beta$-coincident $\gamma$-ray spectrum, in which we have singled out many of the weak peaks listed in
Table~\ref{intensity}.  Those peaks identified with unbracketed energies have measured intensities; those with bracketed energies are only assigned a limit.
The unlabeled strong peak in parts (a) and (b) of the figure is at 1567 keV.  The unlabeled one in part (c) is at 3211 keV; the one in part (d) is at 3848
keV.} 
\end{figure}

\begin{figure}[t]
\centering
\includegraphics[width=0.78\columnwidth]{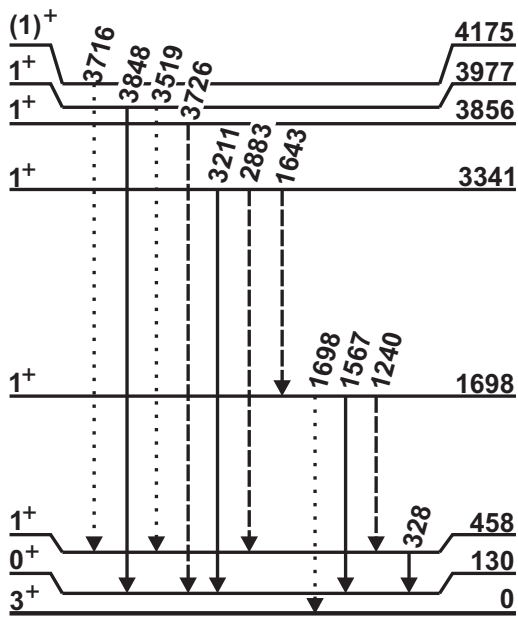}
\caption{\label{fig8}Partial level scheme of $^{38}$K, showing the excited states populated by the $\beta$ decay of $^{38}$Ca, and the $\gamma$ transitions
that occur or may occur following the $\beta$ decay.  The four transitions shown with solid lines are the strongest ones.  They have already appeared
in Fig.~\ref{fig2} and correspond with the $\gamma$-ray peaks identified in Fig.~\ref{fig4}.  The dashed lines identify additional weak observed
transitions; and the dotted lines indicate even weaker transitions, for which we set only upper limits.  Note that the 130-keV state is a 924-ms isomer;
it has a very weak $\gamma$-ray branch to the ground state, which is not shown.} 
\end{figure}

The relative intensities we obtained for the $\beta$-delayed $\gamma$-rays observed in the decay of $^{38}$Ca are listed in Table \ref{intensity}, and
portions of the summed $\beta$-coincident $\gamma$-ray spectrum are shown in Fig.~\ref{fig7}, illustrating the energy regions that encompass many of those
weak transitions.  A complete scheme of the energy levels in $^{38}$K populated by $^{38}$Ca $\beta$ decay is presented in Fig.~\ref{fig8}, where all the
$\gamma$-ray transitions appearing in the table are clearly indicated.  Compared to the simplified scheme in Fig.~\ref{fig2} this decay scheme includes an
additional level weakly populated by $\beta$ decay and a number of inter-level $\gamma$ transitions.  In determining each $\gamma$-ray intensity in
Table~\ref{intensity}, we have incorporated the $\beta$-detector efficiency given in Table~\ref{table1} for the $\beta$ transition that populates the state
from which each originates.  

Also given in Table \ref{intensity} are the results from the only two previously published studies of the relative $\gamma$-ray intensities \cite{Wi80,An96}.
Our sensitivity is similar to the most recent previous measurement, by Anderson {\it et al.}\cite{An96}, but our results show some significant
discrepancies with theirs.  The most egregious is for the intensity of the 2883-keV $\gamma$ transition, which we determine to be an
order of magnitude less than their result.  It seems very likely that the value quoted by Anderson {\it et al.}~is simply a misprint since it does not seem to be
supported by their own $\gamma$-ray spectrum (see Figure 5 in Ref. \cite{An96}), in which the 2883-keV peak is clearly not one-half the intensity of the
nearby peak at 3211 keV as their tabulated intensity would lead one to believe.  There are two other smaller, but still significant, discrepancies between
our results and those of Anderson {\it et al.} for the peaks at 1643 and 3726 keV.  These discrepancies give us good reason to use our results exclusively
in all subsequent determinations of the Gamow-Teller $\beta$-branching ratios from $^{38}$Ca.

\section{Results}
\label{s:results}

\subsection{Gamow-Teller branching ratios}
\label{ss:br}

\begingroup
\squeezetable
\begin{table*}[t]
\begin{center}
\caption{Measured $\beta$-branching ratios to all the states in $^{38}$K populated by the $\beta$ decay of $^{38}$Ca.
\label{branching}}
\vskip 1mm
\begin{ruledtabular}
\begin{tabular}{cccccccc}
& & & & & & &  \\[-3mm]
$E_{x_i}$\footnotemark[1] & $E_{\beta max}$ & Relative & & Relative & Absolute & & Superallowed   \\
(keV) & (keV) & $\beta^+$ branching & $(1+\xi_i)/(1+\xi)$  & ($\beta^+$+ ec) branching & ($\beta^+$+ ec) branching &log $ft$ & $\F t$ value (s)   \\[1mm] 
\hline
& & & & & & &  \\[-2mm]
130.4 & 5590.1 &  & ~0.99970 & & ~0.7728(16) & ~~~$\,$3.4860(10) & 3076.4(72) \\
458.5 & 5262.0 & ~0.1447(30) & ~0.99985 & ~0.1447(30) & 0.0281(6)  & ~4.804(1) &  \\
1697.8 & 4022.7 & ~~~1.0026($^{+17}_{-15}$) & ~1.00092 & ~~~1.0035($^{+17}_{-15}$) & ~0.1948(13) & ~3.426(3) &  \\
3341.2 & 2379.3 & ~0.0166(13) & ~1.00839 & ~0.0167(13) & 0.0032(3) & 4.19(4) &  \\
3856.0 & 1864.5 & 0.0007(3) & 1.0187 & 0.0007(3) & ~0.00014(6) & ~$\,$5.09(19) &  \\
3977.3 & 1743.2 & 0.0050(8) & 1.0233 & 0.0051(8) & 0.0010(2) & 4.11(9) &  \\
\end{tabular}
\end{ruledtabular}
\vspace{-3mm}
\footnotetext[1]{Values taken from Ref.~\cite{Ca08}.}
\end{center}
\end{table*}
\endgroup

In Sections~\ref{ss:eff}, \ref{ss:bsingles} and \ref{ss:Ngamma}, we obtained values for all the quantities on the right-hand side of Eq.~(\ref{eq:branchingratio}).
Using these results we determine that
\begin{equation}
\label{eq:R'}
 R^{\prime}_{1698}=0.1941(13).
\end{equation}  
However, as already mentioned, Eq.~(\ref{eq:branchingratio}) is strictly valid only for $\beta$
transitions that populate a state which de-excites by the emission of a single $\gamma$ ray, and, as we showed in Sec.~\ref{ss:relgamma}, the situation is actually
more complicated, with the 1698-keV state being populated and de-populated by several weak $\gamma$-ray transitions.  In addition, because we measure $\beta$-$\gamma$
coincidences, we see only the effect of positron emission but are blind to the, albeit small, effects of electron capture.  To acknowledge that further small
corrections are required, we use the notation $R^{\prime}$ in Eq.~(\ref{eq:R'}).

We consider first the effect of the weak $\gamma$-ray transitions involving the 1698-keV state.  Two have measured relative intensities: a 1643-keV $\gamma$ ray
populating the state and a 1240-keV $\gamma$ ray de-populating it.  For a third, at 1698 keV, an upper limit is established (see Table~\ref{intensity}).  If we continue
to use the  normalization employed in Table~\ref{intensity}, the total intensity of $\beta^+$ feeding to the 1698-keV state thus becomes $1.0026^{+17}_{-15}$.
This result appears in the third row, third column of Table~\ref{branching}.

To take account of electron capture, we first have to recognize that both the numerator and denominator in Eq.~(\ref{eq:branchingratio}) need to be corrected for
missing electron-capture decays.  Thus our result for $R^{\prime}_{1698}$ must be multiplied by $(1+\xi_{1698})/(1+\xi)$, where $\xi_{1698}$ is the electron-capture-to-positron
ratio for the $\beta$ transition populating the 1698-keV state and $\xi$ is that ratio for the total decay of $^{38}$Ca.  These ratios, $\xi$, are readily calculated as a
function of $E_{\beta max}$, the $\beta$ endpoint energy, and $Z$ (see, for example, Ref.~\cite{NNDC}) and are entirely independent of nuclear structure.  The results
are $\xi_{1698}=0.00196$ and $\xi=0.00103$.  The ratio $(1+\xi_{1698})/(1+\xi)$ appears in the third row, fourth column of Table~\ref{branching}.

Multiplying our result for $R^{\prime}_{1698}$ in Eq.~(\ref{eq:R'}) by the relative intensity of $\beta^+$ feeding to this state and the electron-capture correction ratio, we
find the final branching ratio for the ($\beta^+$+ ec) transition to the 1698-keV state to be
\begin{equation}
\label{eq:R}
 R_{1698}=0.1948(13).
\end{equation}  
This result also appears in the third row, sixth column of Table~\ref{branching}.
  
The same method can be used for the Gamow-Teller transitions to other levels in $^{38}$K, based on the intensities of the $\gamma$ rays that populate and de-populate
those levels.  Maintaining the same normalization to the intensity the 1567-keV $\gamma$ ray (see Table~\ref{intensity}) and with the help of Fig.~\ref{fig8}, which shows
the placement in the decay scheme of all the observed $\gamma$ rays, we obtain the relative $\beta^+$ branching ratios listed in column three of Table~\ref{branching}.  After
multiplying by the electron-capture correction ratios (column four) and the value for $R^{\prime}_{1698}$ in Eq.~(\ref{eq:R'}), we arrive at the final branching ratios for the
four remaining Gamow-Teller transitions from $^{38}$Ca, which are listed in the sixth column of the table.

To obtain the corresponding log $ft$ values for these transitions, we used the energies, $E_{\beta max}$, from the second column of the table, combined with the $^{38}$Ca half-life
taken from the most recent survey of world data for superallowed emitters \cite{Ha15}: viz. $t_{1/2}$ = 443.77(35) ms.  These data were used as input to the log $ft$ calculator
available at the National Nuclear Data Center (NNDC) \cite{NNDC} web site.    The results obtained appear in the seventh column of Table~\ref{branching}.  They range from
3.4 to 5.1, which is well within the range that characterizes allowed $0^+$$\rightarrow$\,$1^+$ transitions \cite{Si98}.

\subsection{Branching ratio for the superallowed transition}

The total branching ratio for all five Gamow-Teller transitions -- to the $1^+$ states at 459, 1698, 3341, 3856 and 3977 keV -- is 0.2272(16).  This is simply the
sum of the corresponding branching ratios in column six of Table~\ref{branching}.  In principle, a smaller uncertainty would result from our summing only the $\gamma$ rays
that populate the 130-keV state or the ground state, because contributions from the weak $1^+$$\rightarrow$\,$1^+$ $\gamma$ rays would cancel out in the sum.  However, the latter are so weak
that the uncertainty turns out to be essentially the same for both methods.

Shell-model calculations described in Sec.~\ref{ss:GT} convincingly rule out the possibility of large numbers of unobserved weak transitions to higher excited states
that sum to appreciable strength -- the Pandemonium effect \cite{Ha77,Ha02} -- so we can safely conclude that the Gamow-Teller sum we have obtained accounts for all
the non-superallowed strength in the decay of $^{38}$Ca. 

The branching ratio for the superallowed $0^+$$\rightarrow$\,$0^+$ transition to the 130-keV analog state is thus 0.7728(16), a result we obtain simply by subtracting
the total Gamow-Teller branching ratio from unity.  Note that, as we do so, we convert the relative precision of our measurement, which is 0.70\% (= 0.0016/0.2272), to a
relative precision of 0.21\% (= 0.0016/0.7728) for the quantity we have sought to obtain: the superallowed branching ratio.  The log $ft$ value for this transition
appears on the top line of column seven in Table~\ref{branching}.  In this case, where the greatest precision is required, we used the full calculation for the statistical
rate function, $f$, as described in Ref.~\cite{Ha15}.

\begin{table}[b]
\caption{\label{ebudget}Uncertainty budget for $^{38}$Ca branching ratios}
\newcommand\T{\rule{0pt}{2.6ex}} 
\newcommand\B{\rule[-1.2ex]{0pt}{0pt}}
\begin{ruledtabular}
\begin{tabular}{ldd}
&
\multicolumn{2}{c}{Uncertainty (\%)} \\
\cline{2-3} 
\textnormal{Source} &
\multicolumn{1}{c}{$\sum$GT \T\B} &
\multicolumn{1}{c}{$0^+$$\rightarrow$\,$0^+$} \\ 
&
\multicolumn{1}{c}{branches} &
\multicolumn{1}{c}{branch} \\ 
\hline
Counting statistics, $\gamma_{1567}$ \& $\beta$ singles \T & 0.49 & 0.14 \\
Contaminant contribution to $\beta$ singles                        & 0.30 & 0.09  \\
$\sum{\gamma}$/$\gamma_{1567}$                               & 0.25 & 0.08  \\
Coincidence summing with 511-keV $\gamma$'s            & 0.21 & 0.06 \\ 
HPGe detector efficiency                       &  0.20 & 0.06 \\
Dead time                                           & 0.07 & 0.021 \\
Bremsstrahlung coincidence summing                     & 0.05 & 0.015 \\
$^{38}$Ca component of $\beta$ singles            & 0.06 & 0.017 \\
Random preemption of real coincidences &  0.04 & 0.012 \\
\\
Total uncertainty & 0.70 & 0.21 \\ 
\end{tabular}
\end{ruledtabular}
\end{table}

\subsection{Uncertainty budget}  

The complete uncertainty budget for our $^{38}$Ca branching-ratio measurement is given in Table \ref{ebudget}, where we present two relative uncertainties (in percent)
for each contribution.  The first is expressed relative to the total Gamow-Teller branches; this can be regarded as the uncertainty as it applies to the
measurement itself.  The second is expressed relative to the superallowed branching ratio, which is the derived quantity of principal interest.

Evidently, the largest contribution to the total uncertainty arises from counting statistics for the 1567-keV $\gamma$ ray and the $\beta$ singles.  Ultimately, the
measurement depended on our detecting the $\gamma$ ray from a 19\% branch of the decay of $^{38}$Ca in the presence of thirty times as many 511-keV positron-annihilation
photons from the decays of both $^{38}$Ca and its daughter $^{38m}$K.  In order to keep dead-time and other corrections to a manageable size, we had to limit our counting
rate, which naturally limited the number of decay $\gamma$-ray events we could collect in a week-long measurement.

The next largest uncertainty arises from contaminants in the collected samples.  It is a measure of the precision we have achieved with this measurement, that the
quantification of impurities, which totaled no more than 0.35\% of each sample (see Sec.~\ref{sss:imp}), was the second largest source of uncertainty.  Following this in
importance, were the uncertainties attributed to the determination of $\gamma$-ray intensities relative to that of the well-determined 1567-keV peak, and the correction
to account for coincidence summing with annihilation radiation.  Both were dependent on our necessarily constrained counting rate.

All four of these contributions to the uncertainty budget depend on conditions that specifically apply to our measurement conditions and can be regarded as statistical.
The remaining five are inherent to our basic equipment and techniques; we regard them to be systematic.  With the exception of the uncertainty associated with our HPGe
detector efficiency, all are very much smaller than the statistical uncertainties.

\section{Discussion}
\subsection{Superallowed decay branch}

Our measurement of the branching ratio for the superallowed $0^+$$\rightarrow$\,$0^+$ $\beta$ transition from $^{38}$Ca is the first ever made.
Several branching-ratio measurements for the Gamow-Teller branches have previously appeared in the literature \cite{Wi80,An96,Ca08}, but all were
normalized to a calculated ratio for the superallowed branch based on the expected constancy of the superallowed $\F t$ values\footnote[1]{Often it is the
$ft$ values that have been assumed to be identical for all such superallowed decays.  In fact, it is the $\F t$ values, which include the important radiative
and isospin-symmetry-breaking corrections, that should be identical.}.  We are now able to test that assumption
for this decay.

Our branching-ratio result has already been published in letter format \cite{Pa14} and has been included in the most recent survey of superallowed
$0^+$$\rightarrow$\,$0^+$ nuclear $\beta$ decays \cite{Ha15}.  There it was combined with world-average $Q_{EC}$ and half-life results to obtain an $ft$
value of 3062.3(68) s.  The relationship between this $ft$ value and the $\F t$ value used to extract $V_{ud}$ is given by
\begin{equation}
\F t \equiv ft(1+\delta_R^{\prime}) (1 + \delta_{NS} - \delta_C)
\label{Ftdef}
\end{equation}
where $\delta_C$ is the isospin-symmetry-breaking correction and the terms $\delta_R^{\prime}$ and $\delta_{NS}$ comprise the transition-dependent part of
the radiative correction, the former being a function only of the electron's energy and the $Z$ of the daughter nucleus, while the latter, like $\delta_C$,
depends in its evaluation on the details of nuclear structure.  Taking the values for these three small correction terms from Table IX in Ref.~\cite{Ha15},
we obtain the result that appears in the last column in Table~\ref{branching}: viz. $\F t$= 3076.4(72) s.  With 0.2\% precision, this result for $^{38}$Ca
is competitive with $\F t$ values for the previously well-known superallowed emitters, almost all of which have branching ratios that are nearly 100\% and thus did not
require such a challenging measurement.  The $\F t$ value for $^{38}$Ca decay is entirely consistent with 3072.24(62) s, the average $\F t$ value for the
other thirteen well-known superallowed emitters \cite{Ha15}.

More significant is the fact that the newly obtained $ft$ value for the superallowed branch from $^{38}$Ca provides us the first chance to study a precisely
measured mirror pair of superallowed transitions: {\it viz.} $^{38}$Ca\,$\rightarrow$$^{38m}$K and $^{38m}$K\,$\rightarrow$$^{38}$Ar.  Accepting the constancy
of $\F t$, the ratio of $ft$ values from a mirror pair relates directly to the calculated correction terms $\delta_R^{\prime}$, $\delta_{NS}$ and $\delta_C$
as follows:  
\begin{equation}
\frac{ft^a}{ft^b} = 1+(\delta^{\prime b}_R-\delta^{\prime a}_R)+(\delta^b_{NS}-\delta^a_{NS})-(\delta^b_C-\delta^a_C)~,
\label{ftratio}
\end{equation}
where superscript ``$a$" denotes the decay $^{38}$Ca\,$\rightarrow$$^{38m}$K and ``$b$" denotes $^{38m}$K\,$\rightarrow$$^{38}$Ar.  As explained in
Ref.~\cite{Pa14}, the crucial advantage offered by Eq.\,(\ref{ftratio}) is that the theoretical uncertainty on a difference term such as $(\delta^b_C
-\delta^a_C)$ is significantly less than the uncertainties on $\delta^b_C$ and $\delta^a_C$ individually.  This means that the experimental $ft$-value ratio
can provide a sensitive and independent test of the veracity of the correction terms, particularly $\delta_C$.

Based on data from the recent review \cite{Ha15}, the ratio for the $A$=38 pair is $ft^a/ft^b$ = 1.0036(22).  This value is consistent with $\delta_C$ values
calculated with Woods-Saxon radial wave functions, which yield a ratio value of 1.0020(4), and inconsistent with $\delta_C$ values calculated with Hartree-Fock
radial wave functions, which yield a ratio value of 0.9998(4).  This outcome was an important factor in the elimination of the latter calculations from the
derivation of $V_{ud}$ from the ensemble of superallowed $ft$ values \cite{Ha15}.

\begin{table*}
\caption{Experimental and theoretical excitation energies and $\beta$-decay branching ratios, $R$, to the daughter $1^+$ states in $^{38}$K.  The theoretical
values were obtained from an $sd$ shell-model calculation with effective interactions USD, USD-A and USD-B.  In calculating the branching ratios, we have used
the experimental energies in the phase-space calculation for the first two $1^+$ states, and the theoretical energy for the third $1^+$ state.
\label{t:Rratios}}
\begin{ruledtabular}
\begin{tabular}{rrdrrdrrdrrd}
&  \multicolumn{2}{c}{Expt} & &
   \multicolumn{2}{c}{USD} & &
   \multicolumn{2}{c}{USD-A} & &
   \multicolumn{2}{c}{USD-B} \\ 
\cline{2-3}
\cline{5-6}
\cline{8-9}
\cline{11-12}
\\[-3mm]
\multicolumn{1}{r}{State} &
\multicolumn{1}{r}{$E_x$(keV)} &
\multicolumn{1}{r}{$R(\%)$} & &
\multicolumn{1}{r}{$E_x$(keV)} &
\multicolumn{1}{r}{$R(\%)$} & &
\multicolumn{1}{r}{$E_x$(keV)} &
\multicolumn{1}{r}{$R(\%)$} & &
\multicolumn{1}{r}{$E_x$(keV)} &
\multicolumn{1}{r}{$R(\%)$} \\[1mm]

\hline
& & & & & & & & & & & \\[-1mm]
$1_1^+,T=0$ & 458 & 2.81 & & 630 & 3.77 & & 710 & 5.57 & & 540 & 2.13 \\
$1_2^+,T=0$ & 1698 & 19.48 & & 1720 & 15.0 & & 1760 & 22.5 & & 1500 & 19.4 \\
$1_3^+,T=0$ & 3341 & 0.32 & & & & & & & & & \\
$1_4^+,T=0$ & 3856 & 0.01 & & & & & & & & & \\
$1_5^+,T=0$ & 3977 & 0.10 & & 4140 & 0.41 & & 4080 & 0.44 & & 4220 & 0.24 \\
\end{tabular}
\end{ruledtabular}
\end{table*}

\subsection{Gamow-Teller branches}
\label{ss:GT}

With the ground state of $^{38}$Ca described predominantly as two holes in a closed-shell $^{40}$Ca, the states in $^{38}$K strongly populated by $\beta$
decay from $^{38}$Ca must also have predominantly two-hole configurations according to the selection rules for allowed Gamow-Teller transitions.  Even so, since
detailed spectroscopic studies of mass-38 nuclei do show evidence for some mixing of four-hole two-particle ($4h$-$2p$) configurations with the two-hole ($2h$) ones, 
we can expect to see a few deviations from a simple two-hole model.  Corroboration for both the predominant behavior and for a few deviations can be found in our
$\beta$-decay results.

In Table~\ref{t:Rratios}, we show the results of $sd$ shell-model calculations for $1^+$ states in $^{38}$K involving only two-hole configurations.  We
use the USD effective interaction of Wildenthal \cite{USD} and two more recent updates, USD-A and USD-B, of Brown and Richter \cite{USDAB}.  In all cases we use a
quenched value for the axial-vector coupling constant, $g_{A,{\rm eff}} = 1.0$, which Brown and Wildenthal \cite{BW85} demonstrated to be appropriate for use
in calculations truncated to just $sd$-shell configurations.

These shell-model calculations identify only three $1^+$ states of two-hole configuration in the lowest 4 MeV of excitation energy in $^{38}$K, compared with five
states identified in our $\beta$-decay measurement.  There is very good correspondence between experiment and theory for the lowest two $1^+$ states, both in
excitation energy and in $\beta$-feeding strength.  For the third, fourth and fifth $1^+$ states observed in our experiment, we can suppose that they actually
involve mixing of $2h$ and $4h$-$2p$ configurations, with only their $2h$ component being sampled in $\beta$ decay.  If we take the sum of the $\beta$ strengths to
these three states and compare that with the calculated strength to the single $2h$-model state at $\sim$4 MeV, the agreement is really quite good.  Overall, the USD-B
effective interaction gives the closest match to the experimental results, although all three calculations do rather well.

For precise $\beta$-decay studies such as the one reported here, it is essential to ensure that no decay strength remains unaccounted for.  In particular, one must rule
out -- or correct for -- low-energy $\beta$ transitions to highly excited states, transitions that may individually be too weak to be observed but could be numerous enough
that their total intensity is of significance \cite{Ha77,Ha02}. How many $1^+$ states are there in $^{38}$K between 4.0 and 6.7 MeV, the $\beta$-decay $Q$-value window,
that could in principle be fed in a Gamow-Teller transitions?   Our $sd$ shell-model calculations yield only two states of two-hole configuration in this energy range: 
a $1^+, T=1$ state at 5.5 MeV excitation and a $1^+,T=0$ state at around 6 MeV excitation.  Both these states have negligible calculated $\beta$ branching -- approximately one
per million decays -- and consequently they play no role in the determination of the superallowed branching ratio.

\vspace{8mm}
\section{Conclusions}

We have described in detail our measurement of branching ratios for the decay of $^{38}$Ca.  The results for the superallowed branch, without experimental
particulars, was published in letter format last year \cite{Pa14}.  In addition to those particulars, here we have included full information on the Gamow-Teller branches
as well.  The latter agree well with shell-model calculations, which is important since the shell model is also used to calculate the isospin-symmetry-breaking
corrections, $\delta_C$.  Agreement with the Gamow-Teller branching ratios and with the $ft$ value ratio in Eq.~\ref{ftratio} lends added credibility to the
approach used in the analysis of $0^+$$\rightarrow$$\,$$0^+$ superallowed decays and the extraction of $V_{ud}$ to test CKM unitarity.

There are three more $T_Z$=-1 superallowed emitters like $^{38}$Ca, which can complete mirror pairs and can currently be produced prolifically enough for
precision $ft$-value measurements: $^{26}$Si, $^{34}$Ar and $^{42}$Ti.  We are now embarked on studies of their decays as well.

\begin{acknowledgments}

This material is based upon work supported by the U.S. Department of Energy, Office of Science, Office of Nuclear Physics, under
Award Number DE-FG03-93ER40773, and by the Welch Foundation under Grant No.\,A-1397.

\end{acknowledgments}

\end{document}